\documentclass[twocolumn,showpacs,preprintnumbers,amsmath,amssymb]{revtex4}

\usepackage{graphicx}
\usepackage{dcolumn}
\usepackage{bm}

\begin{document}
\title{High-temperature cuprate superconductors studied by 
x-ray Compton scattering and positron annihilation spectroscopies}

\author{Bernardo Barbiellini}

\affiliation{Department of Physics, Northeastern University, Boston MA 02115, USA}


\begin{abstract}
The bulk Fermi surface in an overdoped ($x = 0.3$) single crystal 
of La$_{2-x}$Sr$_{x}$CuO$_{4}$ has been observed by using x-ray 
Compton scattering. This momentum density technique also provides
a powerful tool for directly seeing what the dopant Sr atoms are doing
to the electronic structure of La$_2$CuO$_{4}$. 
Because of wave function effects, positron annihilation spectroscopy 
does not yield a strong signature of the Fermi surface in 
extended momentum space, but it can be used to explore the role 
of oxygen defects in the reservoir layers for promoting high temperature superconductivity.
\end{abstract}


\maketitle

\section{Introduction}
Discovered by Bednorz and M\"uller in 1986 \cite{bednorz},
high temperature superconductivity in cuprates 
has mobilized scientists around the 
world, both to understand 
its origin and to find new compounds with 
higher superconducting transition temperature $T_c$.

\begin{figure}
\includegraphics[width=8cm]{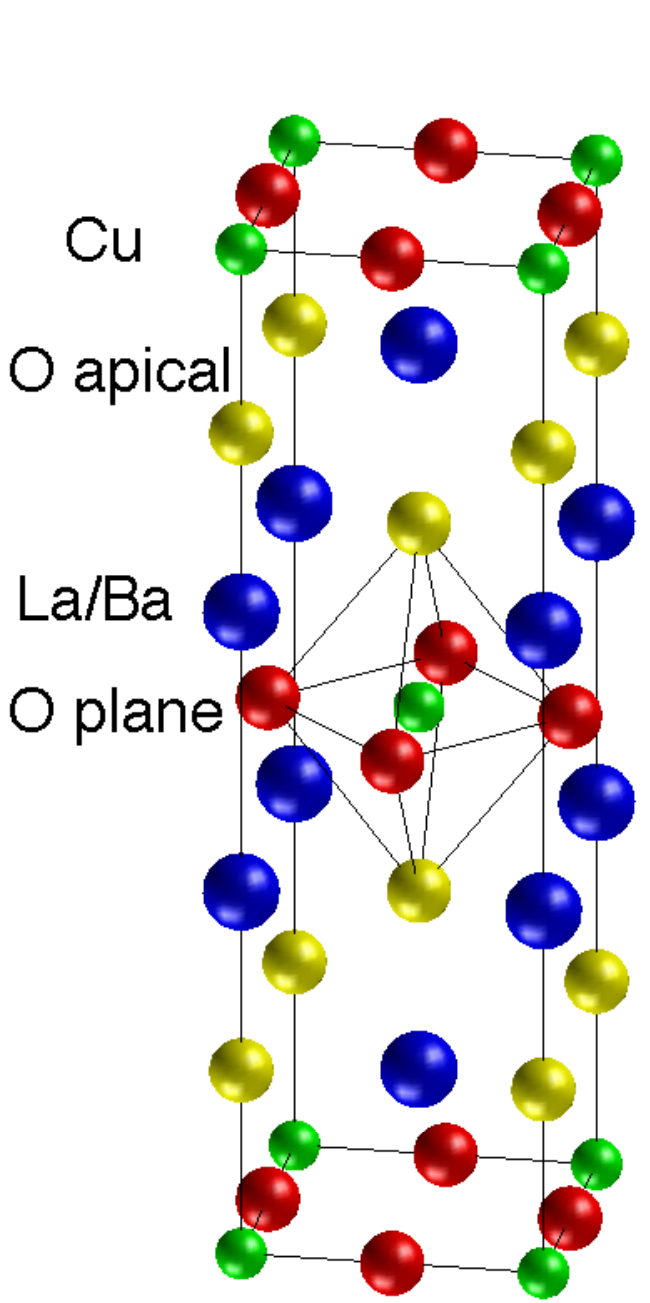}
\caption{\label{fig_STRU} 
The structure of La$_{2-x}$Ba$_{x}$CuO$_{4}$.
The La and Ba atoms (blue spheres) are in 
the charge reservoir layers, which contains 
all the dopant atoms. The so-called apical O atoms
(yellow spheres) are located in the same
layers while the Cu atoms (green spheres) and 
the planar O atoms (red spheres)
belong to the superconducting layers.
The CuO$_6$ octahedron (shown in the middle) 
is formed by a CuO$_4$ plaquette and two apical O 
atoms and determines the important $x^2-y^2$ and $z^2$ 
features of the electronic structures.}
\end{figure}

The lanthanum compounds La$_{2-x}$M$_{x}$CuO$_{4}$ 
have variable impurity content $x$ and M = Ca, Ba, Sr.
Their crystal structure, shown in Fig.~\ref{fig_STRU},
belongs to the K$_2$NiF$_4$ family. 
It is built up from Cu-O superconducting layers 
alternating with rock salt type layers, which provide 
the charge reservoirs.
When $x=0$, these materials have 
antiferromagnetic insulating phases.
It is only with increasing $x$ that the 
metallic state and superconductivity set in.
A naive chemical argument finds
that, when $x=0$, copper has a valency $+2$
and thus contains one hole (missing electron). 
The increase of $x$ creates excess of holes that
causes the oxides to conduct.

However, the physics and chemistry of doping high-temperature 
cuprate superconductors still remain something of a mystery. 
For instance, standard theories cannot readily explain the sudden 
destruction of superconductivity at high doping concentrations. 
For this reason, we have undertaken \cite{prl_bj} 
Density Functional Theory (DFT) calculations 
for nano-sized supercells of La$_2$CuO$_4$  with dopants.
These calculations show that weak ferromagnetism appears 
around clusters of high Ba concentration
and suggest that ferromagnetism and superconductivity 
compete in overdoped samples.
Our studies on doping have also shown how by scattering 
high energy x-rays from single crystals of 
La$_{2-x}$Sr$_{x}$CuO$_{4}$, one can directly image 
the character of holes doped into this material \cite{science}.
At low doping the hole O $2p$ character is much stronger 
than the Cu $3d$ character. However, at high doping the 
Cu $3d$ character becomes dominant. These observations 
are highly significant because via x-ray scattering 
we have new tools for directly seeing 
what the dopant atoms are doing in these materials \cite{platzman}. 
The same advanced x-ray characterization could be used to study 
cathode materials for lithium batteries \cite{jacs}, 
where the octahedron formed by a transition metal
atom and six oxygen atoms plays a key role
like the CuO$_6$ building block in the La$_2$CuO$_4$ 
compound (see Fig.~\ref{fig_STRU}).

Since high $T_c$ superconductivity appears 
to be near a metal-insulator
transition there is a particular
significance attached to the existence of
the Fermi Surface (FS) in the normal state.
Currently, the only direct probe of FS 
is angle-resolved photoemission (ARPES) 
\cite{damascelli}.       
While ARPES has many strengths, this 
has created the problem that  
most of our understanding of cuprates   
is based on measurements from a single,
surface sensitive technique that has been applied       
only to a limited number of materials that cleave.       
Thus, the risk is that experimental artifacts    
can be interpreted as fundamental physics.
Therefore, it is important to develop new approaches
to measuring both electron momentum density and
FS applicable to a much wider class of materials.

\section{2D ACAR spectroscopy} 
The two dimensional angular correlation
of positron annihilation radiation (2D ACAR) 
\cite{acar1,acar2,barnes}
depends, by momentum and energy conservation,
on the two photon momentum density 
$R(p)$ of an annihilating electron-positron pair 
\cite{posreview,bdj}. 
Given a typical electronic momentum
$|p|\sim 10^{-2} mc$, the angle between the two photons
deviates only by a few milliradians from anticollinearity,
since each photon has momentum $\sim mc$ (energy $\sim mc^2$).
Thus angles $\theta_x$ and $\theta_y$ are directly related to
the $p_x=mc\theta_x$ and $p_y=mc\theta_y$ components of $p$.
The two dimensional (2D) ACAR spectrum is measured by $\gamma$
detectors at a large distance from the sample (about 10 meters).
Then one obtains the projection of $R({\bf p})$
\begin{equation}
N(p_x,p_y)=\int dp_z~R({\bf p}).
\end{equation}
Actually, the distribution $N(p_x,p_y)$ is convoluted with 
the resolution function of the experimental setup and the typical
experimental resolution is of the order $0.03$ a.u. of momentum.
The 2D ACAR has been successful for
a determination of the FS in many metallic systems \cite{bdj}, 
but similar studies of the high-$T_c$ oxides have met difficulties
\cite{platzman,barnes}. 
Although the electron-positron momentum density measured in a 
positron-annihilation experiment contains FS signatures, 
the amplitude of such signatures is controlled by the extent to 
which the positron wave function overlaps with the states at the Fermi energy. 
Calculations of the positron density distribution indicates that the positron 
does not probe well the FS contribution of the 
Cu-O planes in La$_{2-x}$Sr$_{x}$CuO$_{4}$ \cite{wael}. 
Indeed, we see little evidence of FS signatures 
in either the computed or the measured 2D ACAR distributions. 

\begin{figure}
\includegraphics[width=9cm]{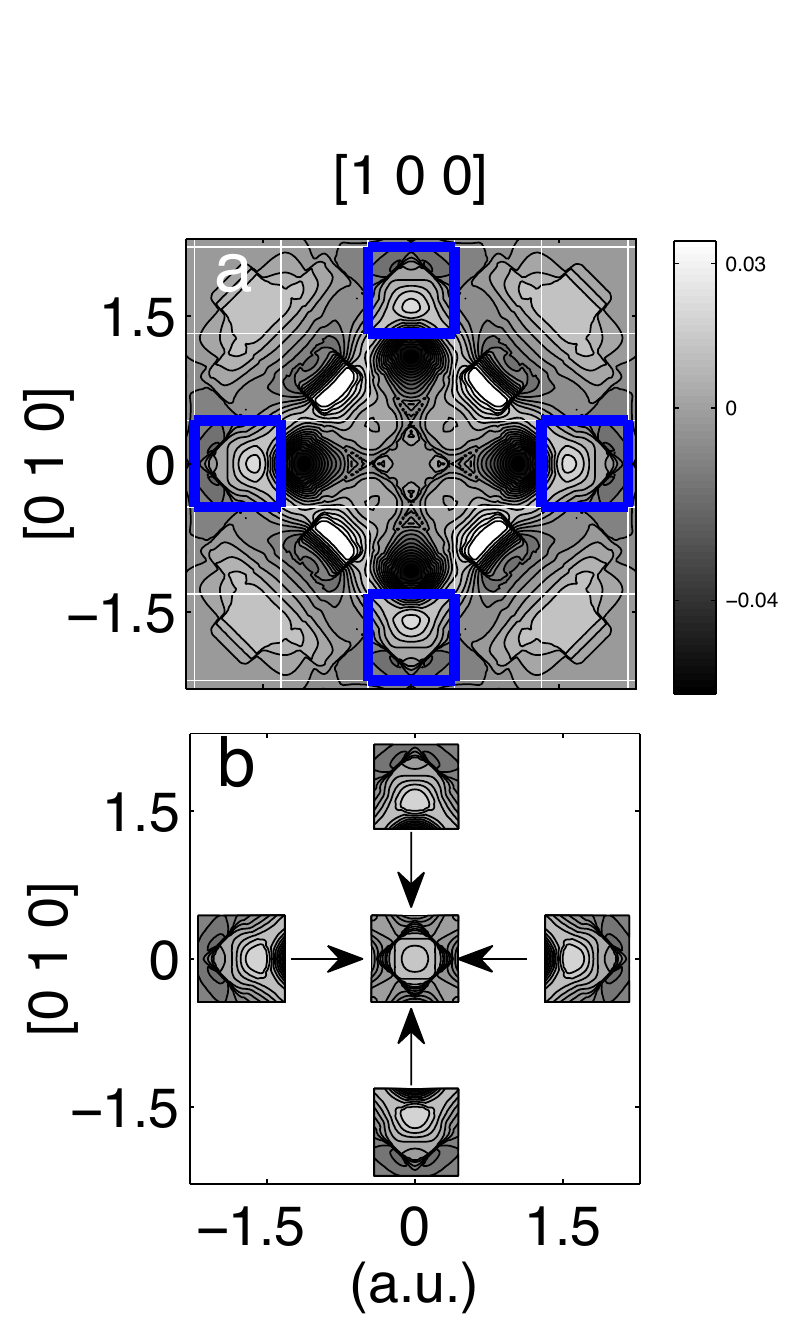}
\caption{\label{fig_EMD} FS analysis from
the 2D electron momentum density distribution:
a) Cylindrical anisotropy for theoretical electron momentum density.
The white lines define the Brillouin zones, while blue squares 
indicate a family of zones where the FS 
features are particularly strong. 
(b) The blue squared regions are isolated from the rest 
of the spectra and folded back to the central region.
The color scale is in units of the 2D momentum density at the origin.
The total amplitude of the 2D momentum density anisotropy is of the 
order of 8\% in excellent agreement with the experiment.}
\end{figure}

\section{X-ray Compton scattering spectroscopy} 
Compton scattering refers to inelastic x-ray scattering 
\cite{PT_phil,ray,piero} in the deeply inelastic 
limit in which the non-resonant scattering cross-section 
is well-known to become proportional to the ground 
state momentum density \cite{kaplan}. 
Compton scattering thus provides a uniquely powerful probe of 
the many body ground state wave function. 
The technique is also well suited for 
investigating disordered alloys because, being a ground state probe, 
it does not require long electron mean free paths. 
By contrast, transport type experiments such as 
the De Haas-van Alphen effect are limited to low defect and impurity concentrations. 
Like other light scattering techniques, 
Compton scattering is genuinely a bulk probe, 
which is not complicated by the presence of surface effects involved in ARPES. 
Historically, the capabilities of the Compton technique have been slow to realize 
in practice because only a limited momentum resolution 
($\sim 0.4$ a.u.) was possible 
with gamma ray sources. However, with the availability 
of second and third generation synchrotron light sources 
and improvements in detector technology,  
it has become possible now to achieve momentum resolution of about $0.1$ a.u. in wide classes of materials, 
and of about $0.01$ a.u. in low-Z systems \cite{PT_phil,cooper}. 
These advances have sparked a renewed interest in Compton scattering as a tool for investigating Fermi liquid and bonding 
related issues and also as a unique window on hitherto inaccessible correlation effects in the electron gas. 
We have developed extensive methodology for obtaining momentum density and highly accurate Compton profiles (CPs) within the 
Fermi liquid framework \cite{bdj,kkr}
and generalized it to the correlated Resonating valence bond (RVB) ground state \cite{agp1,agp2}. 
These studies have shown, for example, the surprising degree to which 
Li deviates from the Fermi liquid paradigm \cite{agp3}. 
These results have also prompted us to look for ways of going beyond the standard DFT framework 
by using antisymmetric geminal products (AGP) of pair wavefunctions to construct the many-body state in a solid. 
The AGP is an important building block of the RVB.
This scheme can be applied to highly correlated electron systems such as high temperature superconductors.
In the ground state of these materials, multi-configuration effects lead to non-integer occupancies 
of one-electron states. The AGP and RVB calculations have shown that the deviation from integer occupancy becomes 
stronger as correlations grow. However,
the important finding of our recent work in an overdoped ($x = 0.3$) single crystal 
of La$_{2-x}$Sr$_{x}$CuO$_{4}$ \cite{wael}
is that strong correlation effects that ordinarily reduce the Compton anisotropy 
amplitudes are no longer effective at this doping.
The quantitative agreement between DFT calculations and momentum density experiments 
(2D ACAR and 2D Compton scattering reconstruction) 
shown in Ref.~\cite{wael} suggests that Fermi liquid 
physics is restored in the overdoped regime.  A FS signal was also clearly observed 
by Compton scattering in the third Brillouin zone along [100] 
and the calculated FS topology was found to be in good accord with the corresponding experimental data. 
In particular, Fig~\ref{fig_EMD}~(a) shows the cylindrical anisotropy of the theoretical 2D electron moment density. 
Because the FS is periodic,  a complete FS must exist in each Brillouin zone, but with its intensity modulated 
by electronic wave functions effects. 
For a predominantly $d$ character at the FS, these wave functions will strongly suppress spectral
weight near the origin, so the FS breaks are most clearly seen in higher Brillouin zones. 
These FS breaks appear superimposed on the momentum
density in the form of discontinuities which can occur in any Brillouin zone. 
In Fig.\ref{fig_EMD}~(a), the Fermi surface breaks are the regions where
the contours run closely together. This means that the electron momentum density varies rapidly at these locations.
Fig.\ref{fig_EMD}~(a) shows the calculated Fermi breaks in several Brillouin zones. In particular, in the third  
zones framed with blue squares, the arc-like features are theoretically predicted FSs associated with Cu-O planes. 
Due to the tetragonal symmetry, a rotation
of the spectrum by $\pi/2$ will generate symmetry-related regions (blue squares) with equivalent strong FS features.
These regions are isolated in Fig.\ref{fig_EMD}~(b).  By folding back
only these regions as shown by the arrows of Fig.\ref{fig_EMD}~(b), we produce a full FS, 
where strong wave functions effects are substantially circumvented.

\section{Oxygen defects in high temperature superconductors}
The positron lifetime is
an ideal tool to detect low concentrations of  
defects in materials \cite{posreview}.
It is determined by the positron-electron contact density, which 
is remarkably higher than the unperturbed electron density \cite{barnes}.
The ratio of these two densities is called the {\em enhancement factor} 
and it is used in ACAR calculations as well.
The agreement between the positron annihilation 
theory and the experiment is usually excellent 
within the Generalized Gradient Approximation (GGA) \cite{gga1,gga2}.
The GGA can also be safely applied for the calculation of 
positron annihilation rates at defects in cuprates \cite{gga3}.

Curiously, it has been difficult to obtain cuprates single crystals of 
a sufficient size and perfection. In fact, some of these substances 
have atomic defects such as oxygen vacancies and interstitials, 
others contain charge density waves and some others present 
themselves with a number of different phases. 
Are all these only accidental complications of minor importance 
to high temperature superconductivity? 
Or, are they on the contrary connected 
to essential properties of high temperature superconductivity?  
An empirical law \cite{zaanen} (only valid for high temperature 
superconductors) suggests that 
\begin{equation}
k_B T_c \tau \sim \hbar,
\label{Eq_planck}
\end{equation}
where $k_B$ is the Boltzmann's constant,
$\hbar$ is the Planck constant, 
$T_c$ is the superconducting 
critical temperature and $\tau$ 
is the electron relaxation time in the normal state.
Therefore, in order to increase $T_c$, one
can increase the electron scattering (i.e. lower $\tau$)
by introducing certain impurities or defects.
The intense level of scattering given 
by Eq.~\ref{Eq_planck} corresponds to 
the so-called  {\em Planckian dissipation limit}, 
beyond which Bloch-wave propagation becomes inhibited, 
i.e. the quasi-particle states themselves become incoherent. 
The fact that the relaxation rate $1/\tau$ scales with $T_c$ 
implies that the interaction causing this anomalous scattering 
is also associated with the superconducting pairing mechanism 
\cite{hussey}.

Interestingly, a different route to hole doping is to 
introduce excess of oxygen in the La$_2$CuO$_4$ compound, 
which is possible because there are extra sites in the reservoir
layers for the oxygen atoms to occupy.
La$_2$CuO$_{4+\delta}$ when doped in the range 
of $0.1 \le \delta \le 0.12$ achieves the best $T_c$, 
however the form of the oxygen interstitials distribution 
\cite{fratini} can change both $T_c$ and $\tau$ dramatically.
So far as the technique of positron spectroscopy is concerned, 
it has become clear that the propensity to oxygen defects \cite{gga3}
has an essential influence on positron annihilation measurements.
For example, we have shown that positrons are trapped
at oxygen vacancies in the superconducting compound  
YBa$_2$Cu$_3$O$_{7-\delta}$ while this trapping
becomes negligible in the non-superconducting sample
where Y has been replaced by Pr \cite{shukla}. 
Our goal is therefore to show via positron 
annihilation spectroscopy that oxygen 
defects (both interstitials and vacancies) 
are connected to essential properties of 
high temperature superconductivity \cite{scojb}. 

\section{Conclusion}
The quantitative agreement between the calculations 
and the experiment for ACAR as well as momentum density anisotropies 
suggests that $x = 0.3$ overdoped La$_{2-x}$Sr$_{x}$CuO$_{4}$ 
can be explained within the conventional Fermi-liquid theory. 
The FS analysis from Compton scattering confirms previous ARPES FS 
measurements showing an electron-like FS in the overdoped regime. 
This validation is important since we provide via deep inelastic x-ray scattering 
experiments a truly bulk-sensitive image of 
momentum density maps of electrons near the Fermi level. 
In general, this momentum density information is difficult to extract 
from ARPES experiments due to difficulties associated 
with matrix element effects and the 
well-known surface sensitivity of ARPES. 
However, improvements in the momentum resolution are still 
needed to bring Compton scattering into the fold of 
mainstream probes for the cuprates FS. 
Higher momentum resolution can also enable
the study of the FS smearing due 
to the superconducting energy gap opening \cite{barnes}.
Positron annihilation spectroscopy cannot probe well 
the FS of the Cu-O layers,
but it can be extremely useful to explain why oxygen defects at
reservoir layers play an important role 
in high temperature superconductivity \cite{scojb}.

\noindent {\em Acknowledgments:}
This work is supported by the USDOE, Grants No. DE- FG02-07ER46352 
and No. DE-FG02-08ER46540 (CMSN) and benefited from the allocation of 
supercomputer time at NERSC and Northeastern Universityâ 
Advanced Scientific Computation Center (ASCC).


\end{document}